\documentclass{emulateapj}

\usepackage{graphicx}
\shorttitle{The mid-IR atmosphere of Betelgeuse}
\shortauthors{Ravi et al.}

\begin{document}

\title{The non-uniform, dynamic atmosphere of Betelgeuse observed at mid-infrared wavelengths}

\author{V. Ravi\altaffilmark{1}, E. H. Wishnow, C. H. Townes, S. Lockwood, H. Mistry, K. Tatebe}
\affil{Space Sciences Laboratory and Department of Physics, University of California, Berkeley, CA 94720, 
USA; vravi@ssl.berkeley.edu}

\altaffiltext{1}{Present address: School of Physics, University of Melbourne, Parkville, VIC 3010, Australia}

\begin{abstract}

We present an interferometric study of the continuum surface of the red supergiant star Betelgeuse at 
11.15\,$\mu$m wavelength, using data obtained with the Berkeley Infrared Spatial Interferometer each year 
between 2006 and 2010. These data allow an investigation of an optically thick layer within 1.4 stellar 
radii of the photosphere. The layer has an optical depth of $\sim$1 at 11.15\,$\mu$m, and varies in 
temperature between 1900\,K and 2800\,K and in outer radius between 1.16 and 1.36 stellar radii. 
Electron-hydrogen atom collisions contribute significantly to the opacity of the layer. The layer has 
a non-uniform intensity distribution that changes between observing epochs. These results indicate that 
large-scale surface convective activity strongly influences the dynamics of the inner atmosphere of Betelgeuse, and 
mass-loss processes.

\end{abstract}

\keywords{infrared: stars --- stars: individual (Betelgeuse) --- circumstellar matter --- stars: mass-loss --- 
techniques: interferometric}

\section{Introduction}

Mass-loss from massive stars in the red supergiant phase is fundamental to the 
chemical enrichment of the interstellar medium \citep{g89}. Red supergiants, 
with initial masses of 10$-$50$M_{\odot}$, are known to lose mass 
at rates of up to 10$^{-4}M_{\odot}$/yr \citep{h86}. Mass-loss rates are 
commonly estimated indirectly, using observations of infrared continuum excesses 
\citep[e.g.,][]{spb89} or measurements of CO rotational line intensities \citep[e.g.,][]{ddd+10}. 
Outflowing material is also directly surveyed through observations of systematic spectral line displacements 
\citep{am35} and interferometric studies over multiple epochs \citep{bdh+96}. The physical 
processes involved in red supergiant mass-loss are however not well understood. While mass-loss 
from less massive stars, such as AGB stars, is understood in terms of significant radial pulsations and 
winds driven by radiation pressure on dust grains, such mechanisms are not applicable to 
red supergiants \citep{jp07}, except those in a superwind phase \citep{yc10}. Rotation in stars that are 
spun-up as they condense from protostellar clouds could play a role in driving mass-loss \citep{hl98}, as could Alfv\`en 
and acoustic waves \citep[e.g.,][]{hm80}. In the case of Betelgeuse ($\alpha$ Orionis, spectral type M2Iab), 
the elevation of cool photospheric gas \citep{lcw+98} and gas at chromospheric temperatures \citep{gd96} has 
been attributed to large-scale surface convective activity \citep{s75,acn84}.

Betelgeuse is perhaps the archetypal red supergiant. Its large angular diameter 
\citep[0.047\arcsec,][]{mp21} and high apparent magnitude ($M_{K}=-4.38$) make it an 
excellent subject for high spatial resolution studies. The continuum angular diameter of 
Betelgeuse has been measured using interferometric techniques at radio \citep{lcw+98}, mid-IR 
\citep[8$-$12\,$\mu$m,][]{wht03a,pvr+07}, near-IR \citep[e.g.,][]{prc+04,hpl+09}, optical 
\citep[e.g.,][]{mp21,bl73,bbw+90} and ultraviolet \citep{gd96} wavelengths, and significant variation 
of the diameter with wavelength is observed. The apparent mid-IR diameter is also observed to vary 
significantly on timescales of years \citep{twh+09}. Numerous interferometric campaigns, from 
optical to mid-IR wavelengths, have further revealed up to three hotspots on the 
stellar surface \citep[e.g.,][]{bbw+90,ybb+00,hpl+09,tcw+07}. These spots range in intensity from a few to a 
few tens of percent of the intensity of the stellar disk. The detection of these spots is generally 
interpreted as direct evidence for the presence of a few giant convection cells on the stellar 
surface \citep[e.g.,][]{chy+10}.

A profusion of circumstellar components have been identified surrounding Betelgeuse. Furthest from the 
star, an analysis of mid-IR spectra and interferometric data by \citet{ddg+94} showed the presence of two 
thin shells of dust at approximately 40 and 80 stellar radii ($R_{*}$), with temperatures of $\sim$400\,K and 
$\sim$200\,K respectively. Later work by \citet{bdh+96} revealed a third, more compact dust shell 
at 4$R_{*}$ that was not apparent in previous data. The dust condensation zone and acceleration 
mechanism are not identified. Radio continuum observations by \citet{lcw+98} at wavelengths between 
0.7\,cm and 6\,cm, probing spatial scales of $2-7R_{*}$, revealed the presence of neutral gas at 
temperatures between 3500\,K and 1300\,K. These temperatures are consistent with the temperature 
profile of the Betelgeuse atmosphere in these regions, as derived from mean temperature and density models 
\citep{hbl01,hrr+09}. At $2R_{*}$, the gas distribution is asymmetric, indicating an anisotropic 
mechanism for elevating the gas above the photosphere. Ultraviolet \citep{gd96} and H$\alpha$ \citep{heh87} 
imaging observations have shown that this cool gas co-exists with asymmetrically-distributed hot chromospheric 
material. Observations of [Fe II] lines by \citet{hrr+09} that correspond to the dominant cool 
atmospheric component showed, surprisingly, that this material was at rest with respect to the star. 
The circumstellar envelope as viewed across the \textit{JHK} wavelength bands was imaged using a `lucky-imaging' 
approach combined with adaptive optics by \citet{kvr+09}, and was found to include a plume extending to 6$R_{*}$.

Closer to the star, strong evidence exists for the presence of a cool molecular layer or shell directly above the 
photosphere within $1.5R_{*}$ \citep[e.g.][]{o04,pvr+07}. This layer is optically thick at mid-IR wavelengths but is 
optically thin in the near-IR. Spectral signatures of water from 
Betelgeuse and the red supergiant $\mu$ Cephei, found using data from the Infrared Space Observatory (ISO), 
were interpreted by \citet{t00a,t00b} as originating from a 1500\,K layer surrounding these 
stars. Modeling of near- and mid-IR interferometric data by \citet{prc+04} confirmed the presence of such a layer, 
with a mid-IR optical depth of $\sim$2 and a temperature of $\sim$2000\,K. Further work by \citet{t06} combining 
both interferometric and spectroscopic data placed the layer surrounding Betelgeuse at beyond $1.3R_{*}$, with a 
temperature of 2250\,K. The layer was shown not to be solely composed of water by \citet{vdv+06}, who suggested 
the presence of amorphous alumina (Al$_{2}$O$_{3}$) to explain the high mid-IR opacity. The spectrum of water lines from 
Betelgeuse is indeed more complex than revealed by the ISO data; signatures of water at photospheric temperatures 
were detected by \citet{js98} and \citet{rhr+06}. Most recently, wide-band, spectrally dispersed mid-IR interferometric 
observations by \citet{pvr+07} were fitted by a layer containing SiO, amorphous alumina and water. High resolution measurements of 
the CO overtone band near 2.3\,$\mu$m \citep{ohb+09} and one-dimensional imaging \citep{owm+11} were modeled 
with a large patch of gas near $1.45R_{*}$ with motions of 10$-$15\,km\,s$^{-1}$ with respect to the star. This emerging model 
of a photosphere and overlying layer for a selection of red supergiant stars allows a better understanding of dust formation and 
red supergiant wind acceleration \citep[e.g.][]{vvh+09}, leading towards a convergent picture of red supergiant 
mass loss.

Studies of the extended Betelgeuse atmosphere form a basis for a better understanding of the atmospheric 
dynamics of red supergiant stars. The existence of a cool layer with high opacity at mid-IR wavelengths directly 
above the photosphere is now established for Betelgeuse. The size, temperature and composition of the layer have 
been investigated, assuming spherical symmetry. Little, however, is known about the spatial structure and 
time-variability of this layer. Consistent instrumentation and modeling procedures have also not been applied 
over multiple observation epochs. In this paper, we present observations of the mid-IR continuum of Betelgeuse 
using the Berkeley Infrared Spatial Interferometer \citep[ISI,][]{hbd+00}. Interferometric data using three 
telescopes were gathered during each year between 2006 and 2010, with angular resolutions sufficient to model 
the stellar shape and potential large-scale features. We describe the observations and image-modeling 
processes in \S2, and perform fits to a photosphere-and-layer model for our observations in \S3. In \S4, 
we discuss the nature of the observed non-uniformities in our model images. We examine the implications of our 
results for models of the Betelgeuse atmosphere and red supergiant mass-loss in \S5, and present 
our conclusions in \S6.


\section{ISI observations of Betelgeuse}

\subsection{Observational methods}

The ISI consists of three 1.65\,m aperture telescopes with a heterodyne detection system at each telescope. 
The double-sideband detection systems are sensitive to $9-12$\,$\mu$m radiation within single sideband widths of 
$\sim$2.7\,GHz. The $^{13}$CO$_{2}$ laser local oscillators can operate at numerous wavelengths corresponding to 
vibrational molecular transitions between wavelengths of 9\,$\mu$m and 12\,$\mu$m. For the observations reported here, 
the local oscillators at each telescope were tuned to a wavelength of 11.15\,$\mu$m, chosen so as to exclude strong 
stellar and telluric spectral lines from the detection band \citep{wht03b}.

\begin{deluxetable}{ccc}
\tabletypesize{\scriptsize}
\tablecaption{Observing log.}
\tablewidth{0pt}
\tablehead{
\colhead{Year} & \colhead{Dates (UT)} & \colhead{11.15\,$\mu$m flux density (10$^{3}$ Jy)}
}
\startdata
2006 & 8, 9, 10 Nov; 7 Dec & 4.2$\pm$0.2 \\
2007 & 14, 15, 16 Nov & 4.1$\pm$0.3 \\
2008 & 22, 23, 24 Sep & 3.8$\pm$0.5 \\
2009 & 4, 5, 7, 9, 18, 20, 22 Nov & 4.4$\pm$0.1 \\
2010 & 5, 17, 18 Nov; 9, 13 Dec & 4.9$\pm$0.3 
\enddata
\end{deluxetable}

\begin{figure*}[h!]
\centering
\includegraphics[scale=0.6]{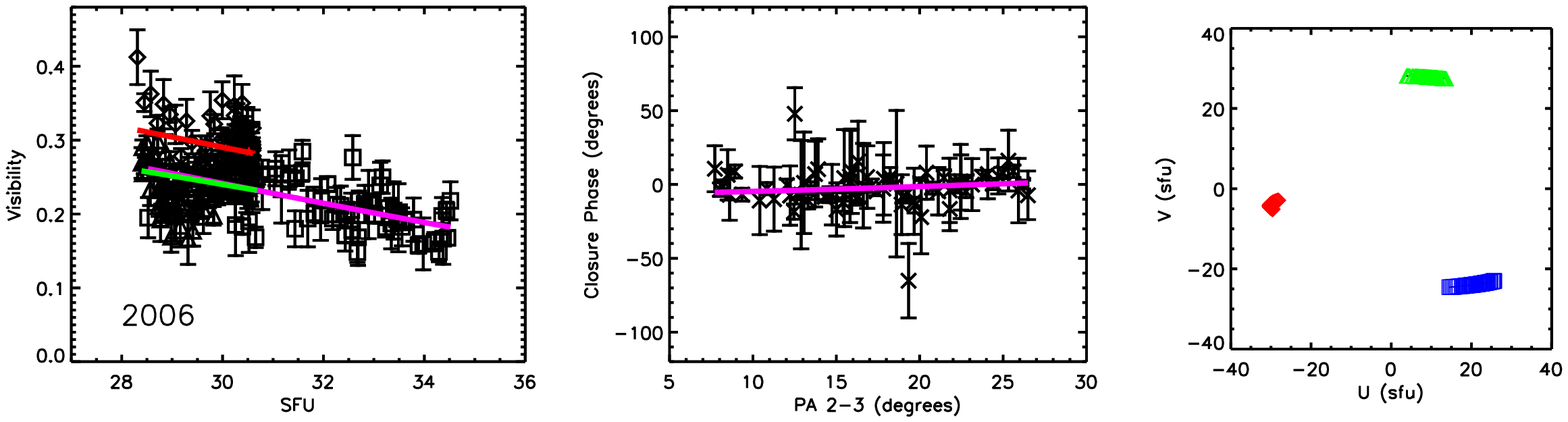}
\includegraphics[scale=0.6]{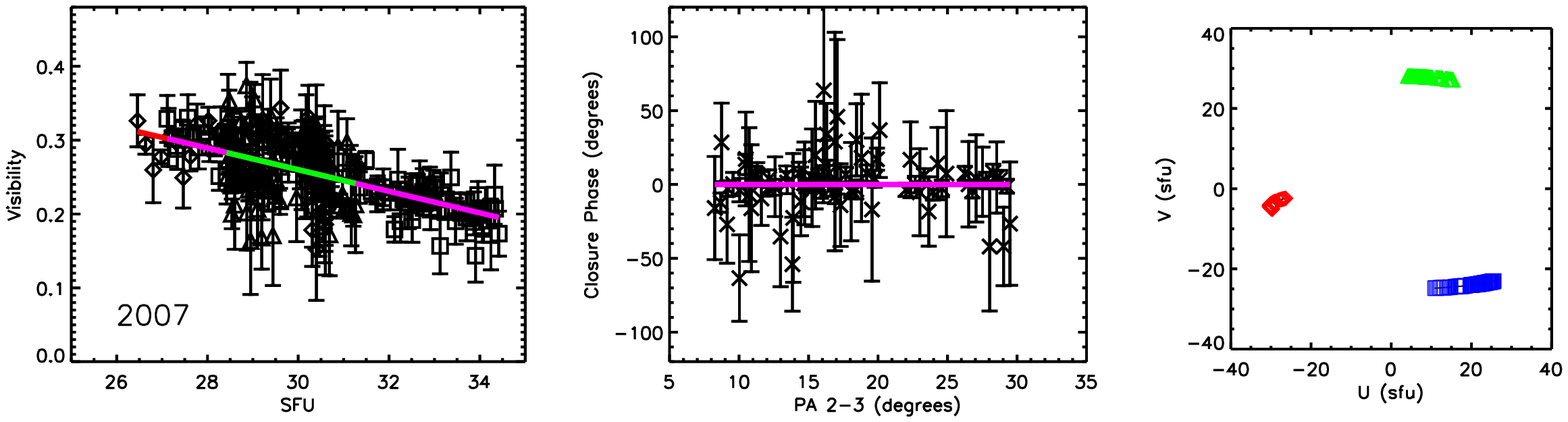}
\includegraphics[scale=0.6]{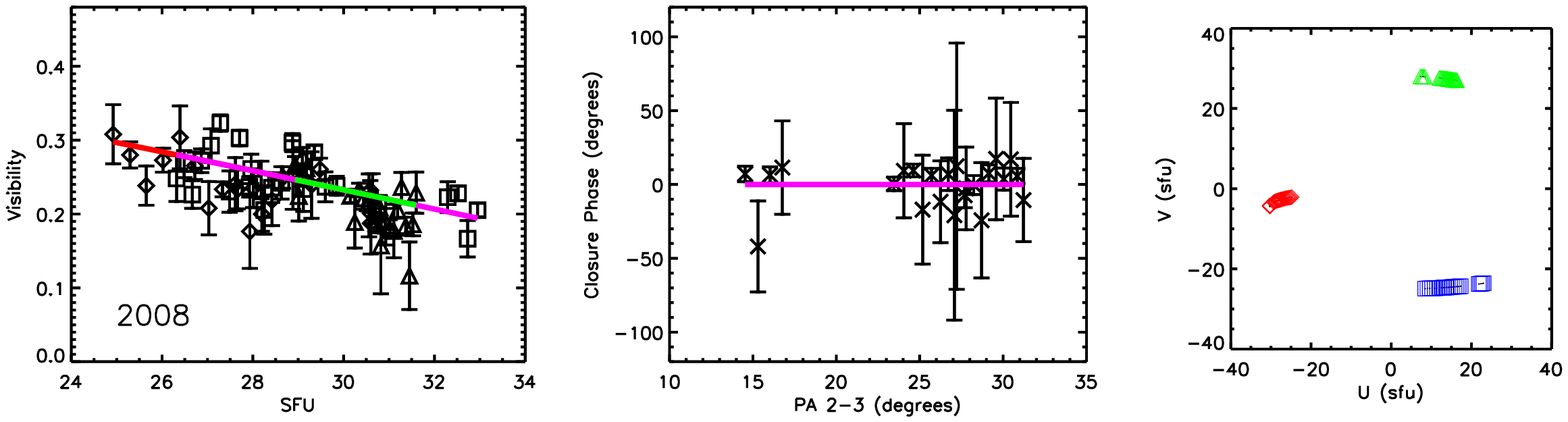}
\includegraphics[scale=0.6]{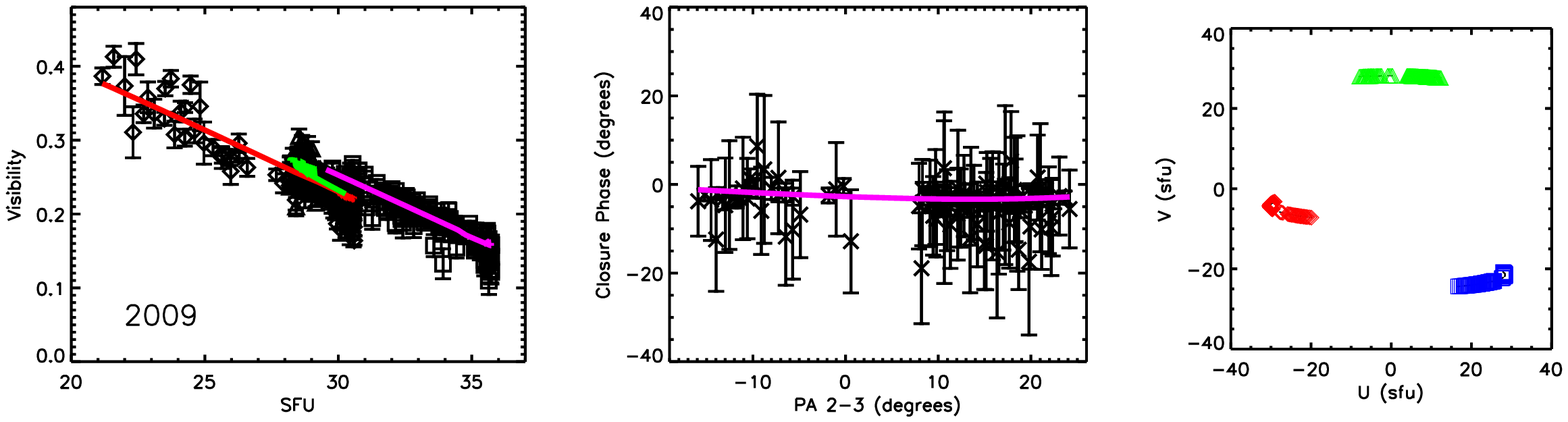}
\includegraphics[scale=0.6]{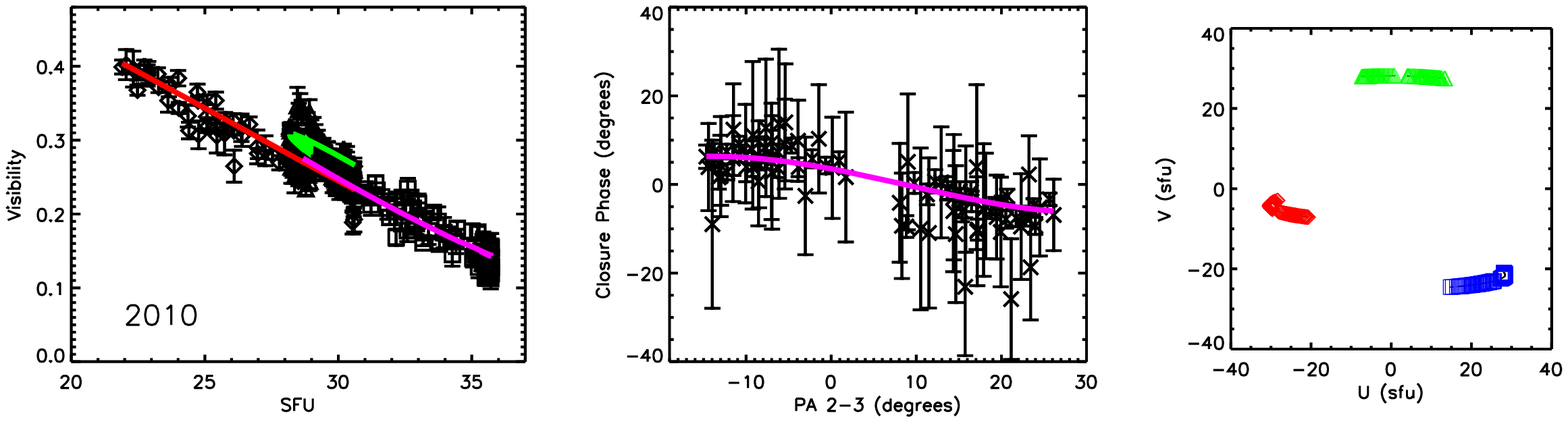}
\caption{Plots of the ISI measurements of Betelgeuse from 2006$-$2010. 
The left panels show visibilities, the center panels show closure phases, and the right panels show UV coverages, for each 
observing epoch (2006 to 2010 from top to bottom). The visibilities for each baseline, 
denoted 1-2, 2-3 and 3-1, are depicted by diamonds, triangles and squares respectively 
(note that the spatial frequency ranges vary from year to year). 
The fitted model visibilities (Table 2) are shown as red, green and magenta lines for baselines 1-2, 2-3 and 3-1 
respectively, and as magenta lines for the closure phases. The closure phases are plotted against the 
position angle, East of North, for baseline 2-3. In the plots of the $UV$ coverage, baseline 
1-2 is represented by red diamonds, baseline 2-3 is represented by green triangles, and baseline 3-1 is represented 
by blue squares.}
\end{figure*}

The van Cittert-Zernicke theorem equates the complex cross-correlation of the electric fields detected at 
two spatially-separated telescopes with the two-dimensional Fourier transform 
of the source brightness distribution within the telescope fields of view. Instead of measuring 
complex cross-correlations, the ISI, like all optical/IR interferometers, 
measures the normalized cross-correlation amplitude, known as the Michelson fringe visibility. The 
Michelson fringe visibility, $V$, is given by 
\begin{equation}
V=\frac{P_{fringe}}{\sqrt{I_{A}I_{B}}},
\end{equation}
where the fringe power, $P_{fringe}$ is the cross-correlation amplitude between signals measured at two telescopes, 
$A$ and $B$, and $I_{A}$ and $I_{B}$ are the source powers measured at telescopes $A$ and $B$ respectively. 
Visibility amplitudes are recorded for each ISI telescope baseline pair. The 
visibility phase in the mid-IR for a given baseline is, however, strongly affected by randomly varying 
atmospheric pathlengths above each telescope, and cannot be recovered from the data. The ISI instead 
measures the closure phase, or the sum of the measured visibility phases for each baseline.  The closure phase 
is independent of optical pathlength differences between pairs of telescopes \citep{j58} and it provides 
information about the point-inversion symmetry of the source. The spatial frequency sampled by the projection 
of a baseline of length $b$, perpendicular to the direction of a source, at a wavelength $\lambda$, is 
proportional to $b/\lambda$. By obtaining measurements at different source hour-angles, multiple spatial 
frequencies at different position angles are sampled.

Visibility and closure phase data were recorded for Betelgeuse over timespans of five weeks or less 
each year between 2006 and 2010 (see Table 1). During this time period, the three ISI 
telescopes were configured in an approximately equilateral triangle with baseline lengths 
between 34\,m and 40\,m. For Betelgeuse, this configuration allows spatial frequencies between 20 
and 37\,SFU (Spatial Frequency Unit; 1\,SFU$=1\times10^{5}$\,cycles\,radian$^{-1}$) to be sampled. For all observations, 
tip-tilt corrections were applied to center the source on \textit{K}-band guider cameras operating at frame rates 
of 27\,Hz. A standard observing sequence consisted of 4.5-minute dataset intervals of staring at the source 
and position-switching between the source and blank fields on either side of the source. The position-switching 
measurements moved the telescopes between the source and the sky every 15\,s. The source powers were measured using 
additional 140\,Hz chopping between the source, or the sky, and cold loads with temperatures near that of the sky. 
A single visibility measurement consisted of either: (1) a fringe power measured during a chopping/position-switching 
dataset, normalized using the source powers during that dataset; or (2) a fringe power measured during a staring 
dataset, normalized by weighted averages of source powers measured during the nearest chopping/position-switching 
datasets. Closure phases were calculated for each triplet of visibility measurements. 

Visibility magnitudes and closure phase zero-point offsets were calibrated using observations of 
Aldebaran ($\alpha$ Taurus) conducted on the same night. Aldebaran was assumed to have a diameter of 
20\,mas and zero closure phase. The visibility calibrations for the two kinds of visibility measurements were 
different because of the different times spent on-source during chopping and staring datasets. 
Data that were affected by systematic errors or poor atmospheric conditions 
were discarded. The total flux densities of Betelgeuse during each observing epoch were also 
estimated using Aldebaran, assumed to have a flux density at 11.15\,$\mu$m of 615\,Jy \citep{mgd98}, 
as a calibrator. These total flux density values are given in Table 1, and represent the total power of Betelgeuse 
within the 5\arcsec$\times$5\arcsec~fields of view of the ISI detectors.

\subsection{Image modeling}

Image models were fitted to the calibrated visibility and closure phase datasets for each 
epoch. The data, analytic fits, and coverage of the $UV$ spatial frequency plane, are shown in 
Figure 1. Effective model-independent image reconstruction was not possible because of the 
sparsity of the $UV$ coverage in all epochs. Our model images 
include a centered uniform disk (UD), and up to two point sources offset from the image centers. The uniform 
disks represent the apparent stellar surface, and the point sources represent localised 
intensity fluctuations in this surface. Emission from the dust surrounding the star at large angular scales, 
while contributing to the total detected intensity, was assumed not to contribute to 
the visibility at the sampled spatial frequencies. The intensity of each image component was fitted 
to a fraction of the total intensity. Other free parameters included the UD 
radius and the positions of the point sources. The model complex visibility at a point $(u,v)$ in 
the $UV$ plane was given by 
\begin{eqnarray*}
V(u,v)&=&\frac{2AJ_{1}(2\pi r\sqrt{u^{2}+v^{2}})}{2\pi r\sqrt{u^{2}+v^{2}}}+ \\
& & P_{1}e^{-2\pi i(ux_{1}+vy_{1})}+P_{2}e^{-2\pi i(ux_{2}+vy_{2})}, 
\end{eqnarray*}
where $A$ is the fraction of the total intensity contributed by the UD, $r$ is the uniform 
disk angular radius, $J_{1}$ denotes a Bessel function of the first kind and of order unity, $P_{n}$ represents 
the fractions of the total intensity of the point sources, and $x_{n}$ and $y_{n}$ represent angular offsets 
to the West and North for the point sources, with $n=\{1,2\}$. 

\begin{deluxetable*}{ccccccc}
\tabletypesize{\scriptsize}
\tablecaption{UD and point source(s) models fitted to visibility and closure phase data.}
\tablewidth{0pt}
\tablehead{
\colhead{Year} & \colhead{UD fraction} & \colhead{UD radius} & 
\colhead{Point source fraction} & \colhead{Pt src $x$ (mas)} & 
\colhead{Pt src $y$ (mas)} & \colhead{Reduced $\chi^2$}
}
\startdata
2006 & 0.51$\pm$0.02 & 24.5$\pm$0.8 & 0.04$\pm$0.01 & -2.4$\pm$0.1 & -23.7$\pm$0.2 & 2.683 \\
2007 & 0.54$\pm$0.01 & 24.8$\pm$0.3 & - & - & - & 2.053 \\
2008 & 0.48$\pm$0.02 & 24.7$\pm$0.8 & - & - & - & 2.416 \\
2009 & 0.557$\pm$0.008 & 26.4$\pm$0.2 & 0.009$\pm$0.001 & -25.0$\pm$0.3 & 10.5$\pm$0.2 & 4.565 \\
 & & & 0.007$\pm$0.001 & 18.3$\pm$0.3 & -20.2$\pm$0.2 & \\
2010 & 0.61$\pm$0.03 & 26.4$\pm$0.3 & 0.025$\pm$0.009 & -29.7$\pm$0.3 & -9.3$\pm$0.1 & 3.65 \\
 & & & 0.009$\pm$0.004 & 8.3$\pm$0.3 & -25.9$\pm$0.1 & 
\enddata
\end{deluxetable*}

We used a weighted least-squares fitting technique for each epoch, with each baseline and the closure 
phase data given equal weights. The differing quality of data from different epochs necessitated using models 
with different numbers of free parameters to fit different epochs. Data from 2006 were 
fitted with a single point source in addition to a UD, data from 2007 and 2008 were only fitted 
with UDs, and data from 2009 and 2010 were each fitted with a UD and two point sources. 
An additional point source was added to the fit for a given epoch only if the reduced $\chi^{2}$ values 
calculated for the full dataset, and for the closure phase data alone, were both decreased upon the addition of the 
point source. We also attempted to fit other image models. These included uniform ellipses, uniform ellipses with point 
sources, and UDs with offset uniform disks or offset two-dimensional Gaussian profiles in place of the point sources. 
None of these trials, however, converged to a satisfactory image, and the angular sizes of the fitted point sources are not determined. 
The fractions of the total intensity contributed by the point sources were not constrained to be positive, as our data are 
sensitive to both bright and dark features on the stellar disk. 

Simulations were conducted to assess possible degeneracies 
in fitted image model parameters given the sparsity of the $UV$ coverage of our datasets, in order to ensure that 
our data were not being over-fitted. No significant degeneracies were found for any epoch. For example, the 2006, 
2009, and 2010 datasets constrain both the locations and brightnesses of the point sources. Closure phase measurements 
at multiple array position angles constrain the position angles of the point sources, and both the closure phase measurements 
and the visibility measurements constrain the angular offsets of the point sources from the image centers.  

The results of the fits, as well as the reduced $\chi^{2}$ for each epoch, are given in Table 2. All the 
fits have reduced $\chi^{2}$ values that are greater than unity. This implies that the data contain 
minor features besides Gaussian noise that are not being fitted by the free parameters of the image 
models. 

ISI data for Betelgeuse obtained during 2006 were also analyzed by \citet{tcw+07}. Despite our choice 
of different observing dates, a different method of calibration and slightly different image fitting 
procedures, our results match those of \citet{tcw+07} within the margins of error. We 
however do not adopt Tatebe et al.'s results for the 2006 epoch in order to maintain the same 
analysis procedure across all epochs.

\subsection{The changing angular diameter of Betelgeuse}

\begin{figure*}[h!]
\centering
\includegraphics[scale=0.75]{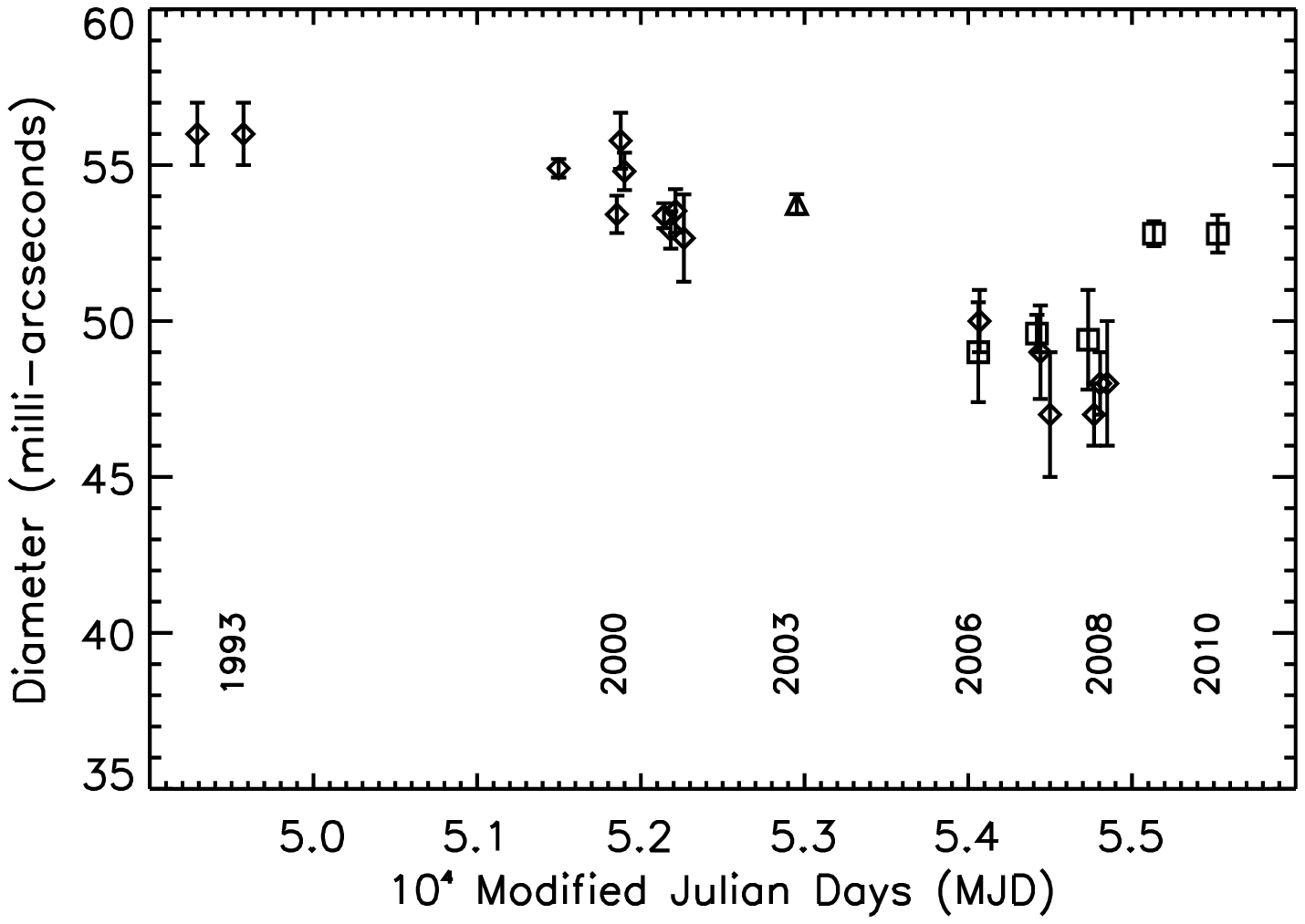}
\caption{The UD diameter of Betelgeuse at 11.15\,$\mu$m measured using the ISI over the period 1993$-$2010. Data between 1993 and 
2009, plotted as diamonds, are taken from \citet{twh+09}, and additional data between 2006 and 2010, plotted 
as squares, are from the present work. The 11.15\,$\mu$m UD diameter measurement of 
\citet{pvr+07}, obtained in 2003 with the VLTI, is plotted as a triangle. The Modified Julian Day (MJD) corresponding 
to a given Julian Day (JD) is MJD$=$JD-2400000.5.}
\end{figure*}

Using ISI observations over the period 1993$-$2009, \citet{twh+09} reported a systematic decrease of 15\% in 
the mid-IR continuum UD diameter of Betelgeuse. Here, we augment this finding with new 
measurements of the UD diameter of Betelgeuse. Figure 2 shows the data plotted in Figure 
1 of \citet{twh+09}, with the new results also included. Though the present work re-analyzes some previous 
observations examined by \citet{tcw+07}, the diameter measurements are obtained from different data sets 
and analyses from those reported by \citet{twh+09}. The measurement of the UD diameter of 
Betelgeuse at 11.15\,$\mu$m by \citet{pvr+07}, obtained using the Very Large Telescope Interferometer 
in 2003, is also plotted. Recently, \citet{owm+11} showed that K-band diameter measurements, made over 
a similar time period, vary less than the mid-IR diameters plotted in Figure 2.

While our results from 2006$-$2008 are consistent to within the margins of error with the trend 
towards a smaller size \citep{twh+09}, the UD diameters of Betelgeuse in 2009 and 2010 are 
significantly larger than those during 2006$-$2008.  
Our analysis method for the 2007 and 2008 datasets essentially matched 
that of \citet{twh+09}, in that only UDs were fitted to the data. Our technique 
however differed somewhat in the methods of calibration and fitting. We consider the variability in 
the apparent size of Betelgeuse in the mid-IR further in \S5.2.

\section{A photosphere-and-layer model for Betelgeuse}

\subsection{The effective temperature}

\begin{figure*}[h!]
\centering
\includegraphics[scale=0.75]{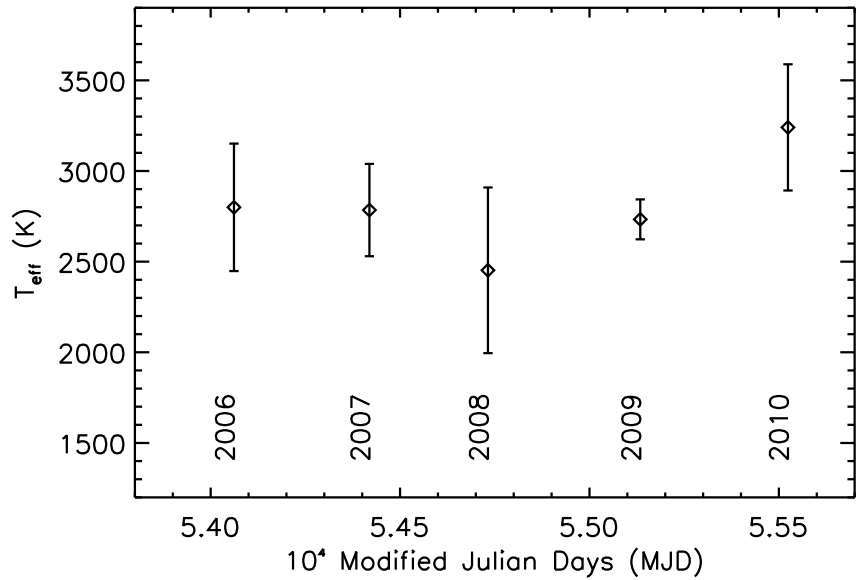}
\caption{The effective temperature of the apparent 11.15\,$\mu$m stellar disk of Betelgeuse for 
each ISI observing epoch in the present work.}
\end{figure*}

Effective temperatures of stellar surfaces are among the fundamental parameters of stars \citep{vab+10}. 
The ISI measurements of the flux density of Betelgeuse, and of the UD sizes and fractions of the 
total intensity, allow the effective temperature of the 11.15\,$\mu$m apparent stellar surface to be 
determined at each observing epoch. We use a Planck function for the power emitted by a blackbody to 
express the effective temperature, $T_{eff}$, as
\begin{equation}
T_{eff}=\frac{h\nu}{k_{B}ln(1+\frac{2h\nu^{3}\Omega}{c^{2}P})}, 
\end{equation}
where $h$ is Planck's constant, $\nu$ is the observing frequency, $k_{B}$ is Boltzmann's constant, 
$\Omega$ is the solid angle subtended by the stellar disk, $c$ is the vacuum 
speed of light, and $P$ is the measured power from the stellar disk in units of W\,m$^{-2}$\,Hz$^{-1}$. 
We do not use the widely-applied technique of calculating effective temperatures using 
estimates for the bolometric flux from the star \citep[e.g.,][]{prc+04,bmc+09} for two reasons: 
radiation at mid-IR wavelengths is strongly affected by thermal emission from circumstellar dust shells 
and bolometric correction factors at these wavelengths are unique to individual stars; and also because the 
correspondance between the 11.15\,$\mu$m scattering surface and the photosphere of Betelgeuse 
is not certain. 

The calculated effective temperatures for the observed stellar disk of Betelgeuse during 
each observing epoch are plotted in Figure 3. We do not account for attenuation by the interstellar 
medium, estimated as modifying the observed flux density by a factor of 0.975 \citep{bdh+96}. We 
also do not account for absorption of the stellar radiation by the dust that surrounds 
Betelgeuse at angular scales of 1\arcsec~or larger, or for the emission from this dust in the 
line of sight to the stellar face. This is because the dust shells are extremely optically thin at mid-IR 
wavelengths when viewed radially, with a total line-of-sight optical depth of approximately 0.0065 \citep{ddg+94}. 

Our measured effective temperatures of the mid-IR surface of Betelgeuse are consistently lower than 
the value of 3641\,K derived by \citet{prc+04} for the photospheric component of a photosphere-and-layer model. 
Furthermore, significant variability is evident between the effective temperatures measured at different 
epochs. We therefore consider sources of mid-IR continuum opacity in the atmosphere of 
Betelgeuse in an attempt to elucidate the scattering surface probed by our observations 
and the cause of the observed variability.

\subsection{The 11.15\,$\mu$m scattering surface of Betelgeuse}

Stellar atmospheres have opacities that vary with wavelength. Stars 
appear smaller at wavelengths where the atmospheric opacity is smaller. The surface 
of a star, when observed at a given wavelength, can be approximated as the location where the 
optical depth at that wavelength, measured from the observer, is approximately unity.  The photosphere 
of a star, however, is the apparent stellar surface observed at a wavelength where the 
opacity matches the Rosseland mean opacity \citep{r24}. The Rosseland mean opacity, $\kappa_{R}$, is given by 
\begin{equation}
\frac{1}{\kappa_{R}}=\left(\int_{0}^{\infty}\frac{\partial B(\nu)}{\partial T}\frac{d\nu}{\kappa(\nu)}\right)
\left(\int_{0}^{\infty}\frac{\partial B(\nu)}{\partial T}d\nu\right)^{-1},
\end{equation}
where $B(\nu)$ is the intensity of radiation from a point in a stellar atmosphere, $T$ is the 
temperature at that point, and $\kappa(\nu)$ is the opacity at a frequency $\nu$. Regions of a stellar 
spectrum with large radiation intensity but low opacity contribute most to determining the Rosseland 
mean opacity. \citet{wht03a} concluded that the mid-IR continuum opacity of 
a gas with solar atomic abundances under surface conditions present on red giant stars closely matches the 
Rosseland mean opacity of such a gas. 

Molecular layers surrounding red giant stars have, however, been shown to significantly 
change the mid-IR opacity from that of a solar abundance atomic gas, while remaining 
relatively transparent in shorter wavelength IR bands \citep{w04,prc+04,vdv+06,wit+09}. 
The \citet{prc+04} model for Betelgeuse, combining  $K$-band visibility data and 
a 11.15\,$\mu$m ISI visibility curve \citep{wdh+00}, includes a shell with a $K$-band 
optical depth of 0.06 and a 11.15\,$\mu$m optical depth of 2.33. \citet{pvr+07} showed 
that this shell was composed of water, amorphous alumina dust and SiO gas, and that the 
11.15\,$\mu$m opacity was dominated by alumina. The presence of alumina in a layer 
surrounding Betelgeuse, within 1.5$R_{*}$, was suggested by \citet{vdv+06} as the only 
dust species that could condense at the temperatures present in this region. 
Alumina dust grains forming close to stars possibly play a role as nucleation sites 
for dust species that condense at lower temperatures \citep[e.g.][]{s77}.

Bremsstrahlung interactions between electrons and neutral hydrogen atoms also contribute to the 
mid-IR continuum opacity in red giant stars \citep{tt06}. This emission mechanism was 
adduced by \citet{rm97} to explain the variation with wavelength of the apparent sizes of the radio photospheres 
of long-period variables, and ascribed to Betelgeuse by \citet{lcw+98}. We discuss this opacity model 
further in \S3.4.

\subsection{Fits to a photosphere-and-layer model}

Prior work, the dramatic variations in apparent mid-IR diameters of Betelgeuse, and the low
effective temperatures all provide clear motivation to model the new ISI observational results
in terms of a photosphere and a molecular layer. Our method is similar to
that of \citet{pvr+07} in that we modeled a spherically symmetric layer surrounding an
opaque stellar photosphere with an assumed diameter of 0.04371\arcsec~and an
assumed surface temperature of 3641\,K \citep{prc+04}.
Visibility and closure phase models for any observed point sources were
subtracted from the data during each observation epoch in order to allow the model to be
spherically symmetric. We did not attempt to fit an asymmetric layer because of the large number
of free parameters involved. The free parameters of our model were: the temperature, $T$, of the 
layer; the optical depth, $\tau$, of the layer along the line of sight to the center
of the star; and, the outer radius, $R$, of the layer.

The layer was assumed to have uniform temperature and opacity, and was also assumed to
have an inner radius that corresponds to the photospheric radius. Radiation within the ISI
band was assumed to arise only from blackbody emission by the layer.

The modeling procedure involved fitting visibilities derived from a model image, created using a 
trial set of free parameters, to the measured visibilities. The measured visibilities for each epoch, 
and their errors, were scaled by the measurements of the measured power emitted by the UD surface of 
Betelgeuse. The flux density of a pixel in the model image grid was calculated as: \\
for $z\leq R_{*}$,
\begin{equation}
I=S(\nu,T_{*})\Omega e^{-\alpha_{S} L(z)}+S(\nu,T)\Omega (1-e^{-\alpha_{S} L(z)}),
\end{equation}
and, for $z>R_{*}$,
\begin{equation}
I=S(\nu,T)\Omega (1-e^{-2\alpha_{S} L(z)}). 
\end{equation}
Here, $L(z)$ describes the optical path length through the layer, $\alpha_{S}$ is the absorption coefficient, 
$z$ is the radial distance of a pixel from the image center, $\Omega$ is the solid angle 
subtended by a single pixel, $S(\nu,T)$ is the specific intensity radiated by a blackbody of 
temperature $T$ at frequency $\nu$, and $T_{*}=3641\,K$ is the assumed temperature of the 
photosphere. $L(z)$ and $\alpha_{S}$ are given by
\begin{eqnarray}
L(z)&=&\sqrt{R^{2}-z^{2}}-\sqrt{R_{*}^{2}-z^{2}},\,z\leq R_{*} \\
L(z)&=&\sqrt{R^{2}-z^{2}},\,z> R_{*} \\
\alpha_{S}&=&\tau/(R-R_{*}). 
\end{eqnarray}
The pixel size of the model images was $1.2\times1.2$ milli-arcseconds. For each iteration of the fit, a 
model image was generated and integrated along one dimension, and a one-dimensional Fourier 
transform was performed in order to generate a set of visibilities. Each model visibility was 
compared to measured visibilities that were closest to it (i.e., within $\pm0.5\,SFU$) in spatial frequency. A 
Levenberg-Marquardt fitting algorithm was employed to calculate the free parameter set 
that minimized the global $\chi^{2}$.  

The results of the fit for each observing epoch are given in Table 3. Errors for each 
parameter were calculated using the resulting covariance matrix. In order to confirm the validity of 
our analyses, we added the visibility and closure phase models of the point sources, that were earlier subtracted 
from the visibility data from each epoch, to the best-fit model visibilities. $\chi^{2}$ values relating 
this sum to the original visibility data were calculated for each epoch, and were found to be 
comparable to the $\chi^{2}$ values of the fits to UDs and point sources shown in Table 2.

The changes with time of the fitted layer temperatures and outer radii are similar to the 
changes with time in the effective stellar temperatures and the apparent UD radii respectively. 
All our fitted layer radii and temperatures are roughly consistent with the 1.3$-$1.5$R_{*}$ 
(0.057\arcsec~to 0.065\arcsec diameter) and 1500$-$2000\,K layers modeled by most authors 
\citep{t00a,o04,t06,vdv+06,pvr+07}. The values we calculate for the 11.15\,$\mu$m optical depth of $\sim$1 
are half of those resulting from the models of \citet{prc+04} and \citet{pvr+07}.  

\begin{deluxetable}{cccc}
\tabletypesize{\scriptsize}
\tablecaption{Temperature, optical depth and outer radius of a layer fitted to visibility and flux density measurements.}
\tablewidth{0pt}
\tablehead{
\colhead{Year} & \colhead{$T$ (K$\times10^{3}$)} & \colhead{$\tau$} & \colhead{$R$ (mas)}
}
\startdata
2006 & 1.9$\pm$0.7 & 0.8$\pm$0.1 & 28.6$\pm$0.7 \\
2007 & 2.5$\pm$0.9 & 0.9$\pm$0.1 & 25.5$\pm$0.9 \\
2008 & 2$\pm$1 & 1.2$\pm$0.2 & 25$\pm$1 \\
2009 & 2.3$\pm$0.8 & 0.99$\pm$0.06 & 26.8$\pm$0.8 \\
2010 & 2.8$\pm$0.6 & 1.05$\pm$0.03 & 29.8$\pm$0.6 
\enddata
\end{deluxetable}

\subsection{Modeling the layer opacity}

The lack of wideband mid-IR spectral information during our observing epochs limits 
modeling of molecular gas and dust species that contribute to the layer opacity. 
We are able, however, to investigate some basic physical characteristics of the layer 
using the opacity mechanism described by \citet{tt06}. This work showed that, 
for solar abundance gas at temperatures similar to those in red giant atmospheres, collision 
processes between neutral hydrogen atoms and 
free electrons dominate the mid-IR opacity. For temperatures of 2500\,K to 3000\,K, 
the primary sources of free electrons were found to be thermally ionized Na and Al. 
The abundances of H$^{-}$, H$_{2}$ and H$^{+}$ relative to H were found to be too low to 
contribute to the opacity through collisions with electrons, and photoionization and 
photodissociation processes were also found to be insignificant. The absorption coefficient due to 
electron-hydrogen atom collisions, first derived by \citet{dl66}, is given in simplified form as 
\begin{equation}
\alpha_{e}=\frac{k_{B}Tn_{e}n_{H}A}{\nu^{2}}
\end{equation}
where $n_{e}$ is the electron number density, $n_{H}$ is the hydrogen atom number density, 
and $A$ is expressed as a third-order polynomial in $T$ \citep{tt06}. This opacity 
mechanism is used to explain the sizes of the radio photospheres of red giant stars \citep{rm97,lcw+98}, 
at different wavelengths. For example, \citet{lcw+98} found that the radio photosphere of Betelgeuse 
appeared larger, yet cooler, for longer wavelengths.

We consider the implications of this mechanism dominating the mid-IR continuum opacity. Equation 9 
can be used to calculate hydrogen atom number densities from the fitted layer optical depths. The electron number 
densities were obtained from Table 2 of \citet{tt06}, which gives the fractional ionizations, relative to $n_{H}$, of 
various metals at different temperatures. We further used the hydrogen number densities 
to calculate the total layer mass, $M_{layer}$, for each epoch, using a distance of $197\pm45$\,pc to 
Betelgeuse \citep{hbg08}. Our results are given in Table 4.

We find masses for the layer of between $3\times10^{-4}M_{\odot}$ and $6\times10^{-4}M_{\odot}$. 
Our values for the hydrogen number density of between $3.5\times10^{11}$\,cm$^{-3}$ 
and $9\times10^{11}$\,cm$^{-3}$, using a hydrogen atomic mass of $1.67\times10^{-24}$\,g, 
are equivalent to mass densities of $\sim10^{-13}$\,g\,cm$^{-3}$. Such values are 
consistent with the hydrogen density derived for the Betelgeuse layer modeled by 
\citet{pvr+07}, where they derived their hydrogen density by assuming a ratio between 
the H$_{2}$O number density and the hydrogen number density of $n_{H_{2}O}/n_{H}\sim10^{-7}$ 
\citep{js98} and using their calculated values of $n_{H_{2}O}$ for the layer. Our values for the 
hydrogen density are therefore also consistent with the values for the water density derived 
by \citet{pvr+07}, though they result from very different methods of calculation.

\begin{deluxetable}{ccc}
\tabletypesize{\scriptsize}
\tablecaption{Number density and mass of the layer.}
\tablewidth{0pt}
\tablehead{
\colhead{Year} & \colhead{$n_{H}$ (cm$^{-3}\times10^{11}$)} & \colhead{$M_{layer}$ ($M_{\odot}\times10^{-4}$)}
}
\startdata
2006 & 5$\pm$1 & 6$\pm$1 \\
2007 & 4.6$\pm$0.9 & 3$\pm$1 \\
2008 & 9$\pm$2 & 4$\pm$2 \\
2009 & 4.3$\pm$0.8 & 3$\pm$1 \\
2010 & 3.5$\pm$0.4 & 5.0$\pm$0.9 
\enddata
\end{deluxetable}

\section{The origin of the point sources in the mid-IR}

\begin{figure*}[h!]
\centering
\includegraphics[scale=0.75]{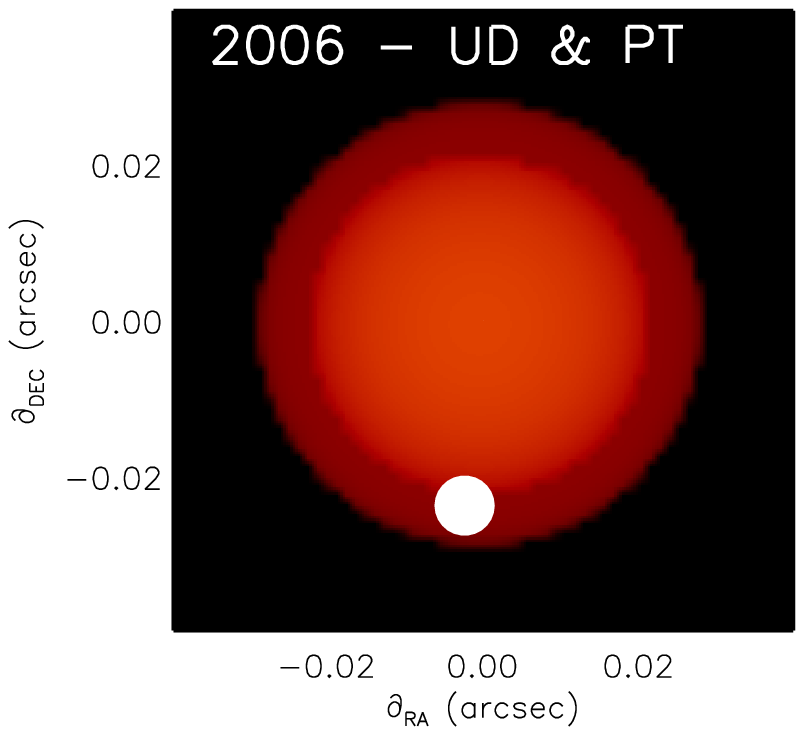}
\includegraphics[scale=0.75]{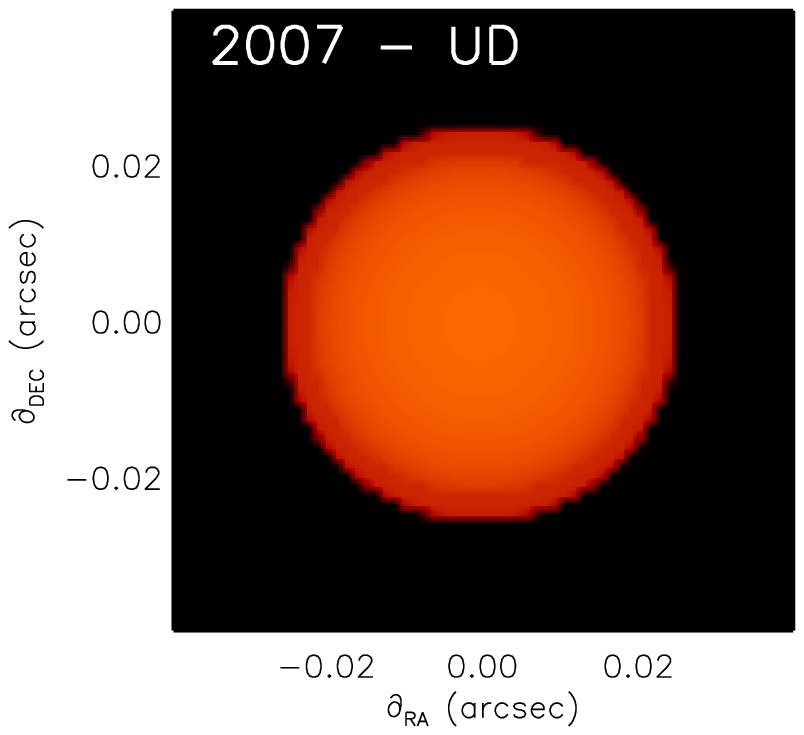}
\includegraphics[scale=0.75]{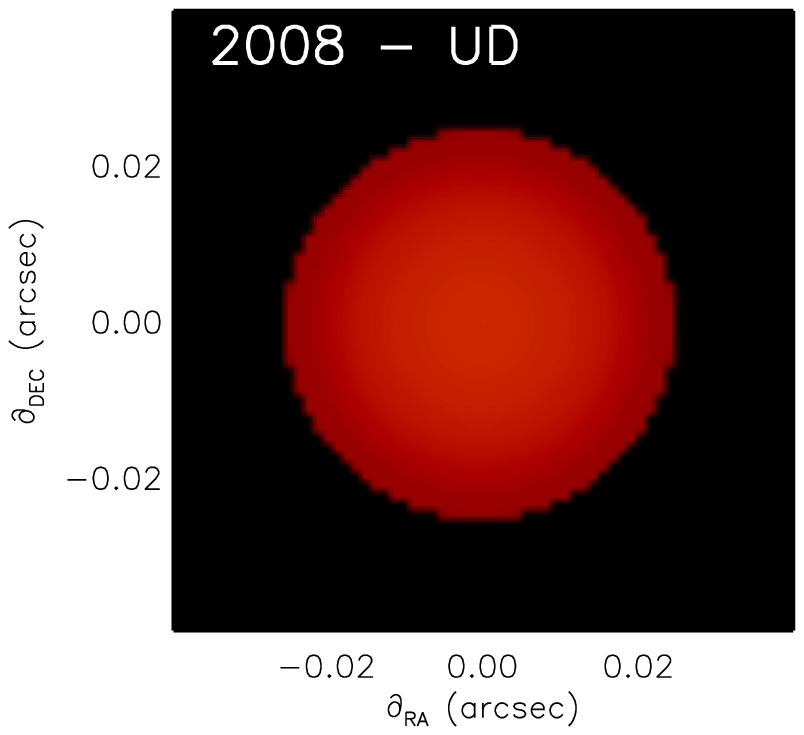}
\includegraphics[scale=0.75]{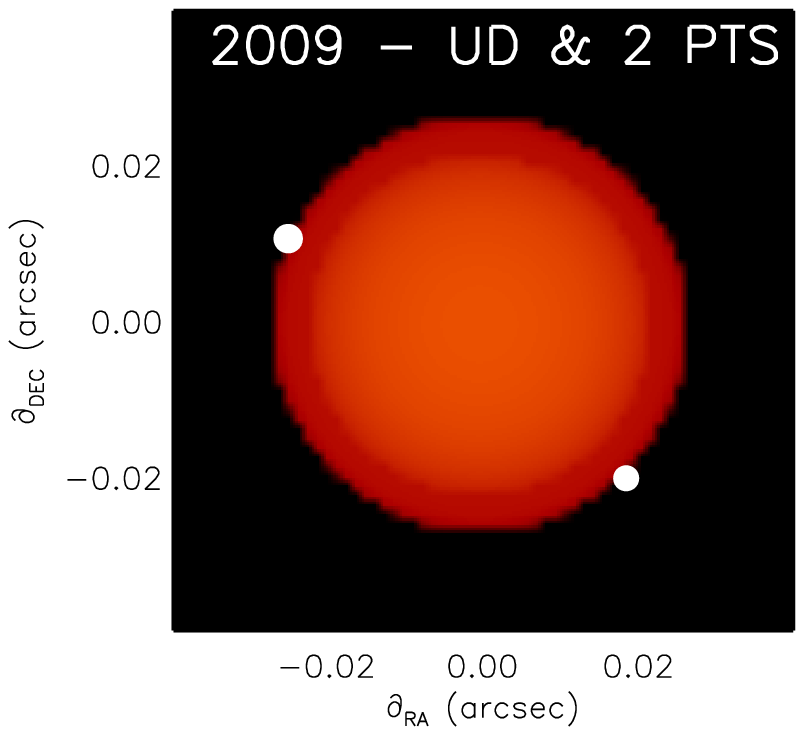}
\includegraphics[scale=0.75]{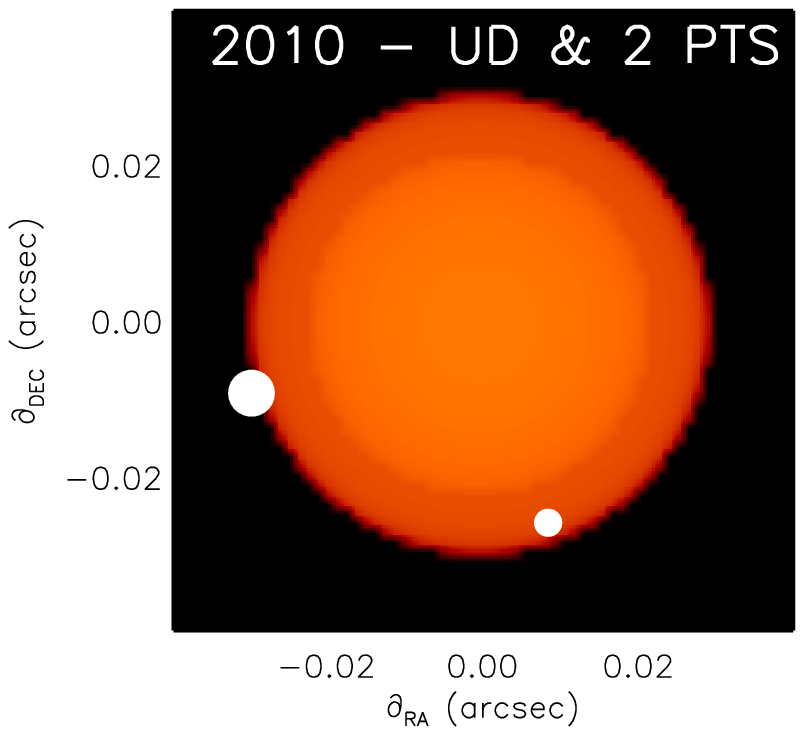}
\includegraphics[scale=0.75]{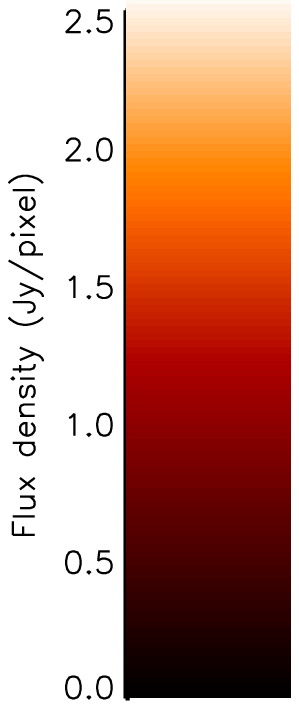}
\caption{Representations of the photosphere-and-layer model intensity distributions fitted to ISI 
observations of Betelgeuse for each observing epoch in the present work. The locations of the fitted point sources are 
shown as white disks: the areas of the disks scale linearly with their intensities. The visibility model fitted to each dataset is 
specified at the top of each image: the 2006 data were fit with a UD and single point source, the 2007 and 2008 data were 
only fit with a UD, and the 2009 and 2010 data were fit with a UD and two point sources. The color bar shows the 
flux density per pixel for each color depicted, and each pixel is 1.2$\times$1.2 milli-arcseconds in size.}
\end{figure*}

We now consider a physical interpretation of the point sources used to model the ISI interferometric 
data on Betelgeuse. Figure 4 shows images of the intensity distribution of Betelgeuse during each observing epoch, calculated 
from the photosphere and layer models, with the point source locations indicated by filled white 
disks. The color brightness of the images is consistent with the same flux density scale, also depicted in Figure 4, 
and the areas of the disks representing the point sources indicate their flux densities calculated from their fitted 
fractions of the total intensity. All point sources are positive in intensity with respect to the star. 

Non-uniformities on the observed stellar disk of Betelgeuse at optical and near-IR wavelengths have been 
reported by several authors \citep[e.g.,][]{bbw+90,hpl+09}. These asymmetries are generally 
modeled either as point sources, or as UDs or two-dimensional Gaussian profiles of fixed, unresolved 
size. The intensities of these `spots' are generally found to be between 5\% and 20\% of the 
intensity of the stellar disk. The distribution and strengths of the spots are found to 
vary significantly across optical and near-IR bands during contemporaneous observations \citep{ybb+00}, 
with the star sometimes appearing featureless in one band and highly non-uniform in another. The 
spots are also found to vary on timescales of less than 8 weeks \citep{wdh97}. One such spot was 
recently resolved in the \textit{H}-band by \citet{hpl+09}, who reported a diameter of 0.011\arcsec 
for this spot. 

These results are consistent with a prediction by \citet{s75} for extremely large-scale convection 
on red giants. Red supergiants are suggested to have only a few tens of convection 
cells on their surfaces, as opposed to the thousands present on the surface of the Sun. Similar predictions 
also result from three-dimensional numerical simulations \citep[e.g.,][]{fsd02,cpj+09}, and an 
initial attempt at matching simulations to interferometric observations was made by \citet{chy+10}. 
\citet{s75} predicted temperature fluctuations of up to $\pm1000$\,K with respect to 
the photospheric temperatures of red giant stars. This is consistent with the intensities of spots observed 
in the visible and near-IR.

The point sources fitted to our mid-IR data do not arise directly from large-scale convective features on the 
surface of Betelgeuse. If we assume a 0.01\arcsec~diameter 
for the weakest of our point sources (see Table 2), which contributes 0.7\% of the 4400\,Jy 
flux density of Betelgeuse in 2009, its temperature of $\sim$660\,K above the effective 
surface temperature, calculated using Equation 2, is consistent with a photospheric hotspot of 
$\sim$1000\,K behind the $\tau=1$ layer. If we assign a similar diameter to any of the other point 
sources we observe, even greater temperatures would result for associated photospheric hotspots. 
The mid-IR point sources are hence more intense than expected from surface temperature fluctuations caused by giant 
convection cells \citep{s75}. Furthermore, every one of our point sources is located beyond the edge of the 
photospheric disk (see Figure 4).

It is clear therefore that the observed asymmetry of Betelgeuse in the mid-IR is caused by a 
non-uniform intensity distribution in the layer. We do not have sufficient sampling of 
the $UV$ plane in any epoch to identify whether the layer is in fact spherically symmetric, but with 
temperature fluctuations, or whether the layer is extended in particular directions. The point sources 
can also be explained in terms of opacity inhomogeneities in the layer. In the 
Rayleigh-Jeans approximation to the Planck law, the effective radiation temperature, $T_{B}$, 
of a column of material with optical depth $\tau$ and temperature 
$T$ is given by 
\begin{equation}
T_{B}=T(1-e^{-\tau}).
\end{equation}
For example, increasing the value of $\tau$ from 1 to 3 increases $T_{B}$ by a factor of 1.5. 
Non-uniform spatial, temperature and opacity distributions of the Betelgeuse layer could 
all explain the observed mid-IR non-uniformities.

\section{Discussion}

\subsection{The extended atmosphere of Betelgeuse}

Molecular layers directly surrounding the photospheres of red supergiant stars, and in 
particular Betelgeuse, have been modeled by a variety of authors under differing 
assumptions for the layer geometry and the dominant opacity source. A layer of water 
with a column density of 10$^{20}$\,cm$^{-2}$ at a temperature of $\sim$1500\,K was 
first suggested by \citet{t00a} for Betelgeuse, based on modeling of near-IR spectra. 
A combination of spectral and interferometric modeling by \citet{o04}, 
assuming a layer geometry identical to ours, suggested a layer temperature of 2050\,K and 
an outer radius of 1.45$R_{*}$, while modeling the mid-IR opacity as being dominated by 
a continuum of water emission and absorption lines. Under the assumption of a layer with 
a well-defined inner radius, \citet{t06} modeled a shell beyond 1.3$R_{*}$ with a temperature 
of 2250\,K, again dominated by water opacity. More advanced spectral modeling by \citet{vdv+06} 
revealed that water could not play a dominant role in determining the mid-IR continuum opacity, and 
instead suggested the presence of amorphous alumina in the molecular layer. This model 
was consolidated by \citet{pvr+07} to suggest a shell between 1.32$R_{*}$ and 1.42$R_{*}$ 
at a temperature of 1520\,K. 

The properties we derive for a layer surrounding Betelgeuse, summarised in Table 3 for 
each observing epoch, are largely consistent with these results. We find temperatures 
that vary between 1900\,K and 2800\,K and outer radii that vary between 1.14$R_{*}$ and 1.36$R_{*}$. 
Our modeling technique most closely matched that of \citet{prc+04} 
in assuming thermal continuum emission from the layer within our observing band. This assumption 
is justified by the featureless spectrum of Betelgeuse within, and around, our observing band
and by the lack of any significant water lines within and surrounding our band \citep{wht03b}. 
The results are consistent with previous models of the Betelgeuse layer.

A consistent picture of the composition of the layer has, however, not been achieved. While 
spectral modeling has uncovered the presence of numerous molecular species, including 
H$_{2}$O, CO, SiO and possibly CN, as well as alumina dust \citep{t06,pvr+07,kvr+09}, the 
opacity of the layer as a function of wavelength is not yet fully determined. Our 
results for the optical depth of the layer of $\sim$1 at 11.15\,$\mu$m wavelength 
are not in agreement with those of \citet{prc+04} and \citet{pvr+07}, who obtain optical depths of 
$\sim$2. 

The number density of atomic hydrogen, based on the 11.15\,$\mu$m opacity arising from electron-hydrogen atom 
collisions, is $\sim$10$^{11}$\,cm$^{-3}$.  This density, given the derived layer size, corresponds to 
H$_{2}$O column densities of 10$^{19}$\,cm$^{-2}$; this result
is consistent with H$_{2}$O column densities measured by \citet{js98} and those derived from spectral 
modeling \citep[e.g.,][]{t06}. We therefore suggest that electron-hydrogen
atom collisions play a significant role in determining the mid-IR continuum opacity of the Betelgeuse 
layer, and that this mechanism must also be accounted for in models for the layer composition.

We have shown that the layer of material surrounding the photosphere of Betelgeuse 
has a non-uniform intensity distribution. A similar result was recently reported by \citet{owm+11} 
based on interferometric measurements of the one-dimensional intensity and velocity profile of CO gas in 
the inner atmosphere of Betelgeuse. \citet{owm+11} also reported a change in the one-dimensional 
CO intensity and velocity profile between two epochs spaced by a year. We find that the intensity 
distribution of the layer varies between each the five observing epochs we present, which are also 
spaced by approximately one year. 

The non-uniformities that we observe in the layer reflect, on much smaller spatial scales, the non-uniformities 
observed in various components of the circumstellar environment of Betelgeuse. A gaseous component at 
similar temperatures to the Betelgeuse layer was shown to have an asymmetric intensity distribution by \citet{lcw+98}. Our 
observations and those of \citet{lcw+98} could be linked by the opacities in both wavelength bands being partly 
caused by electron-hydrogen atom collisions. The opacity due to this mechanism 
is proportional to $\nu^{-2}$, implying that the observed stellar size decreases with shorter 
observing wavelengths. Material at chromospheric temperatures is also observed to 
be asymmetrically distributed \citep{heh87,gd96}, and observations of a circumstellar envelope by 
\citet{kvr+09} at near-IR wavelengths also reveal significant asymmetry. Asymmetry in the 
large-scale ($\sim$1\arcsec) dust distribution surrounding Betelgeuse was also observed by 
\citet{bdt85}.

Molecular layer models have been applied to the few red supergiant stars that have been 
observed with sufficient spatial resolution, such as $\mu$ Cephei \citep{prv+05,t06} and 
$\alpha$ Herculis \citep{o04}. Similar layers are identified around AGB stars 
\citep{wht03a,obd+05}. The properties of individual red supergiant layers vary significantly 
from star to star, and a greater sample of measurements is required to understand what 
determines individual layer properties.

\subsection{Implications of the results reported for red supergiant mass-loss}

The processes by which red supergiant stars form and accelerate dust, while crucial to 
mass-loss scenarios, are little understood. Existing models of stellar atmospheres 
\citep[e.g.,][]{gee+08,p10} cannot predict locations for dust condensation, and attempts to 
model dust acceleration through radiation pressure have also so far been unsuccessful 
\citep[e.g.,][]{w07}. Our multi-epoch study of a layer directly surrounding Betelgeuse 
provides new insight into the processes involved in red supergiant mass-loss.

\begin{figure}
\centering
\includegraphics[scale=0.54]{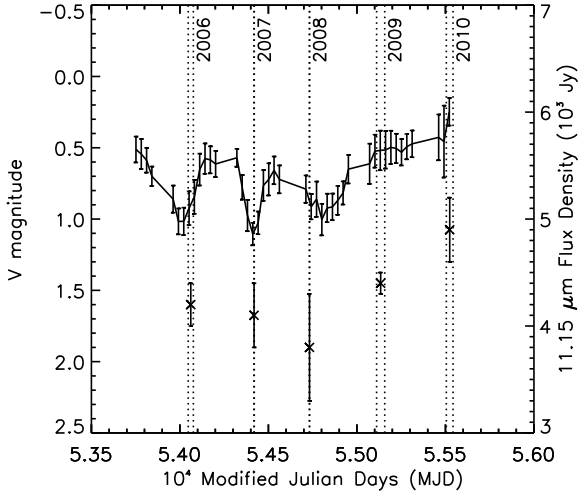}
\caption{Plot of the visual magnitude, and total 11.15\,$\mu$m flux density, of Betelgeuse over the period 2006$-$2010. The 
visual magnitude measurements, joined by a solid line, are 30-day averages of data obtained from the AAVSO online archive. 
The 11.15\,$\mu$m flux density measurements, also listed in Table 1, are shown here as crosses. The timespans of each ISI 
observing epoch are bounded by vertical dotted lines.}
\end{figure}

We have shown that the basic properties of the layer, given in Table 3, exhibit striking 
variability from year to year. The layer is found to decrease in size and increase in 
optical depth over the years $2006-2008$, and increase in size and decrease in optical depth 
between $2009-2010$. The layer temperature does not show such systematic variability, 
although the largest layer size (in 2010) does correspond to the highest temperature.
The variability in the apparent mid-IR size of Betelgeuse reported by \citet{twh+09} can 
also be interpreted to indicate variations in the layers. Intriguingly, a similar phenomenon 
was reported by \citet{p22}, who noted a decrease in the size of Betelgeuse at visible wavelengths 
followed by an increase. 

Figure 5 shows measurements of the visual magnitude of Betelgeuse during between 2006 and 2010, as 
well as the ISI measurements of the total 11.15$\mu$m flux density (also listed in Table 1). The 
timespans of the ISI observing epochs are also plotted. The visual magnitude measurements were obtained from the 
online database\footnote{http://www.aavso.org/data-download} of the American Association of Variable 
Star Observers (AAVSO). The data, collected by a multitude of volunteer observers, were 
averaged over 30-day timespans; each timespan with less than 10 measurements was discarded. 
The visual variability of Betelgeuse over our observing epochs was approximately 0.9 magnitudes. 
The visual intensity of Betelgeuse is not likely to represent that of the
photosphere because the apparent stellar diameter at visible wavelengths is greater than that 
at near-IR wavelengths \citep{ybb+00}. Also, using contemporaneous interferometric observations of 
Betelgeuse at different wavelengths, \citet{ybb+00} found that hotspots evident in 
the visible were not present in the near-IR. Observations of Betelgeuse at visible wavelengths 
could represent regions within the layer observed at mid-IR wavelengths. It is interesting that 
the mid-IR flux density of Betelgeuse, as well as the effective temperatures plotted in Figure 3, 
and the visual magnitude appear somewhat correlated, particularly in their increase during the 2009 and 2010 
epochs. Previous authors \citep{ksb06} have concluded that the visual variability of Betelgeuse is probably 
indicative of quasi-periodic pulsations, and is too great to be explained by variations in the hotspot 
distribution.

The layer of material surrounding Betelgeuse is a candidate location for the onset of dust formation. A number of 
authors have identified the presence of alumina dust directly above the photosphere 
\citep{vdv+06,pvr+07,vvh+09}. The derived physical parameters of the layer, such as the temperatures and hydrogen 
mass densities, are suitable for the formation of amorphous alumina \citep{w06,dss08}. While 
some aspects of the modeled variability of the layer might be due to variations in 
the photosphere, measurements of changes in the apparent mid-IR appearance of Betelgeuse 
must be largely caused by variations in the layer. If alumina contributes 
significantly to the mid-IR continuum opacity, these variations would represent variations in the 
dust content of the layer. Our estimates of changes in the mass of the layer, on the 
order of $10^{-4}M_{\odot}$ (see Table 4), might indicate evolution in the layer. 

The point sources we observe in the layer of material surrounding Betelgeuse and their changes suggest that the layer 
dynamics are anisotropic. Asymmetries observed in the circumstellar environment of 
Betelgeuse are generally linked to the presence of giant convection cells on the stellar 
surface. Such convection cells are thought to shape the circumstellar environment of Betelgeuse 
by elevating cool photospheric material \citep{lcw+98} and driving large-scale chromospheric 
motions \citep{gd96}. Our results indicate that giant convection cells have a significant role in 
shaping the Betelgeuse layer. This is because the mechanism that elevates the layer material above 
the photosphere is clearly anisotropic. The large intensities of the observed point-source components imply the 
action of large-scale photospheric features, and their variability between observing epochs 
demonstrates the transient nature of the cause of the point sources. Changes in the distribution 
of convection cells on the photosphere of Betelgeuse could also influence the overall layer 
properties. While the nature of the non-uniform brightness distribution of the layer is yet to be 
determined, our observations represent evidence for the role of giant convection cells 
in shaping the inner atmosphere of Betelgeuse.

\subsection{Future work}

Observations of a sample of red supergiant stars with high spatial resolution and wide bandwidth, 
over multiple epochs, are necessary to form a general picture for the role of molecular layers in mass-loss 
processes. More detailed interferometric imaging of Betelgeuse at mid-IR wavelengths is also 
required to understand the true role of giant convection cells in shaping the layer. Simultaneous 
visible, near-IR, and mid-IR interferometry could also provide interesting information on correspondences between 
different regions of the layer. Interferometric studies with both high spatial resolution 
and high spectral resolution are planned for the ISI \citep{wmr+10}, and have been performed at 
near-IR wavelengths by \citet{ohb+09} and \citet{owm+11}. Such investigations are aimed at revealing the molecular 
compositions, kinematic structures and true extents of red supergiant layers. More detailed modeling of 
layer opacities is also required, taking into account contributions 
from molecular absorption and emission, dust and electron-hydrogen atom collisions. Upcoming optical and 
infrared interferometer systems, such as the Magdalena Ridge Observatory Interferometer and the MATISSE 
instrument at the Very Large Telescope Interferometer, as well as high spatial resolution observations 
with the next generation of extremely large telescopes could finally clarify our understanding of red 
supergiant mass loss.

\section{Conclusions}

The ISI has been used to conduct observations of Betelgeuse over 2006$-$2010 in the mid-IR continuum with high 
spatial resolution. By fitting visibility and closure phase measurements with image models, the apparent stellar size 
was measured at each observing epoch, and up to two point sources were sometimes detected at the edge of the 
stellar disk. The results were interpreted in terms of an absorbing layer surrounding an assumed 3641\,K photosphere. 
We have shown that:
\begin{enumerate}
\item Observations during every epoch can be modeled in terms of a layer surrounding the photosphere, with 
optical depths of $\sim$1 at 11.15\,$\mu$m wavelength, outer sizes between 1.16$R_{*}$ and 1.36$R_{*}$, and 
temperatures between 1900\,K and 2800\,K.
\item Electron-hydrogen atom collisions contribute to the opacity of the layer.
\item Assuming a non-varying photosphere, the layer exhibits significant variability.
\item The layer has a non-uniform intensity distribution that also varies from year to year.
\end{enumerate}
Both the non-uniformity and the variability of the modeled layer suggest that giant convection cells 
play an important role in shaping the inner atmosphere and the mass-loss dynamics of Betelgeuse.

\acknowledgments

W. Fitelson, B. Walp, C.S. Ryan, D.D.S. Hale, A.A. Chandler, K. Reichl, R.L. Griffith, 
V. Toy, as well as many undergraduate researchers all participated in these observations, 
and their excellent help is greatly appreciated. This research made use of the SIMBAD database. 
Fitting programs used the mpfit IDL routines of C. B. Markwardt. We are grateful for support from the 
Gordon and Betty Moore Foundation, the Office of Naval Research, and the National Science Foundation.

{\it Facilities:} \facility{ISI}

\end{document}